\documentclass[twocolumn,superscriptaddress,showpacs,amsmath,amssymb,showkeys,aps,prc]{revtex4}
\usepackage{dcolumn}
\usepackage{bm}
\usepackage{graphicx}
\usepackage[dvips]{color}
\usepackage{epsfig}
\usepackage{float}
\usepackage{hyperref}
\usepackage{subfig}
\usepackage{verbatim}
\usepackage{dcolumn}% Align table columns on decimal point
\usepackage{setspace}
\usepackage{rotating} %Need this for sidewaystable								 

\newcommand{\melissa}{M\textsc{elissa}}

\begin{document}

\title{Low-background gamma counting at the Kimballton Underground Research Facility}

\author{P.~Finnerty} 
\email[Corresponding author, E-mail: ]{paddy@unc.edu}
\affiliation{Department of Physics and Astronomy, University of North Carolina, Chapel Hill, NC, USA}
\affiliation{Triangle Universities Nuclear Laboratory, Durham, NC, USA}
\author{S.~MacMullin}
\affiliation{Department of Physics and Astronomy, University of North Carolina, Chapel Hill, NC, USA}
\affiliation{Triangle Universities Nuclear Laboratory, Durham, NC, USA}
\author{H.O.~Back}
\affiliation{Department of Physics, North Carolina State University, Raleigh, NC, USA}
\affiliation{Triangle Universities Nuclear Laboratory, Durham, NC, USA}
\author{R.~Henning} 
\affiliation{Department of Physics and Astronomy, University of North Carolina, Chapel Hill, NC, USA}
\affiliation{Triangle Universities Nuclear Laboratory, Durham, NC, USA}
\author{A.~Long}
\affiliation{Department of Physics and Astronomy, University of North Carolina, Chapel Hill, NC, USA}
\author{K.T.~Macon}
\affiliation{Department of Physics and Astronomy, University of North Carolina, Chapel Hill, NC, USA}
\author{J.~Strain}
\affiliation{Department of Physics and Astronomy, University of North Carolina, Chapel Hill, NC, USA}
\affiliation{Triangle Universities Nuclear Laboratory, Durham, NC, USA}
\author{R.M.~Lindstrom}
\affiliation{National Institute of Standards and Technology, Gaithersburg, MD, USA}
\author{R.B.~Vogelaar}
\affiliation{Department of Physics, Virginia Polytechnic Institute and State University, Blacksburg, VA, USA}


\date{\today}

\begin{abstract}
The next generation of low-background physics experiments will require the use of materials with unprecedented radio-purity. A gamma-counting facility at the Kimballton Underground Research Facility (KURF) has been commissioned to perform initial screening of materials for radioactivity primarily from nuclides in the $^{238}$U and $^{232}$Th decay chains, $^{40}$K and cosmic-ray induced isotopes.  The facility consists of two commercial low-background high purity germanium (HPGe) detectors.  A continuum background reduction better than a factor of ~10 was achieved by going underground.  This paper describes the facility, detector systems, analysis techniques and selected assay results.
\end{abstract}

\maketitle

%_______________________________________________ACTUAL TEXT___________________________________________
\section{Introduction}
\begin{figure*}[htp]
	  \centering
      \includegraphics[width=0.8\textwidth]{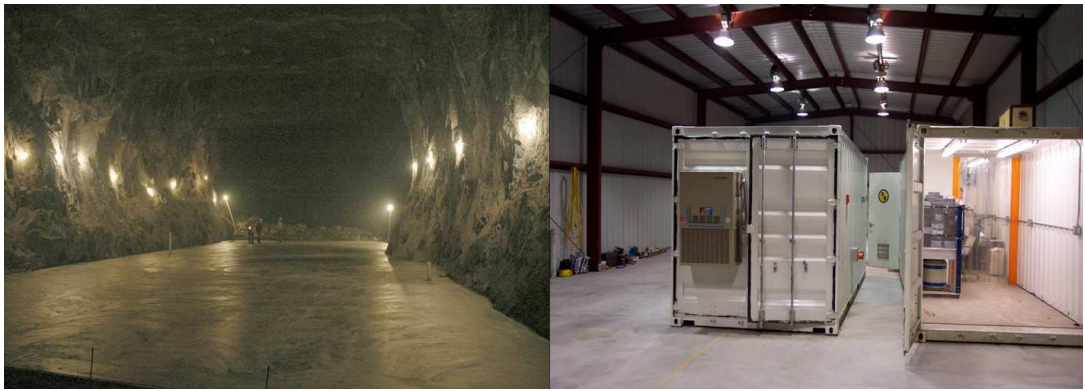}
\caption{KURF before and after construction of the laboratory building.  Modified shipping containers (MSCs) are shown in the right panel (Color online).} \label{fig:beforeafter}
\end{figure*}

We have commissioned a low-background gamma-counting facility at the Kimballton Underground Research Facility (KURF). KURF is located at Lhoist North America's Kimballton mine in Ripplemead, Virginia. The experimental hall is located on the mine's $14^{th}$ level at a depth of 1450 m.w.e (meters of water equivalent shielding).  The overburden consists of 520 m of dolomite, limestone and other sedimentary rock.  Experiments are housed in a 30 m $\times$ 11 m laboratory building that was completed in October 2007 (Fig.~\ref{fig:beforeafter}). The laboratory's general infrastructure is maintained primarily by collaborators at Virginia Polytechnic Institute and State University. KURF has an office, air filtration, power, water, phone and ethernet. KURF also has the advantage of drive in access, making it simple to transport personnel and equipment to the experimental hall. Liquid nitrogen (LN$_2$) used in the laboratory is stored in a 2.4 m$^3$ portable dewar that can be transported to the surface to be refilled as necessary.  Radon levels in the laboratory have been found to vary from 37 Bq/m$^3$ in the winter to 122 Bq/m$^3$ in the summer.  The detectors used in this paper are housed in sealed modified shipping containers (MSCs) within the laboratory building (Fig.~\ref{fig:beforeafter}).

\section{Detectors \& Shielding}
\begin{figure}[htpb]
	\centering
	\includegraphics[width=0.5\textwidth]{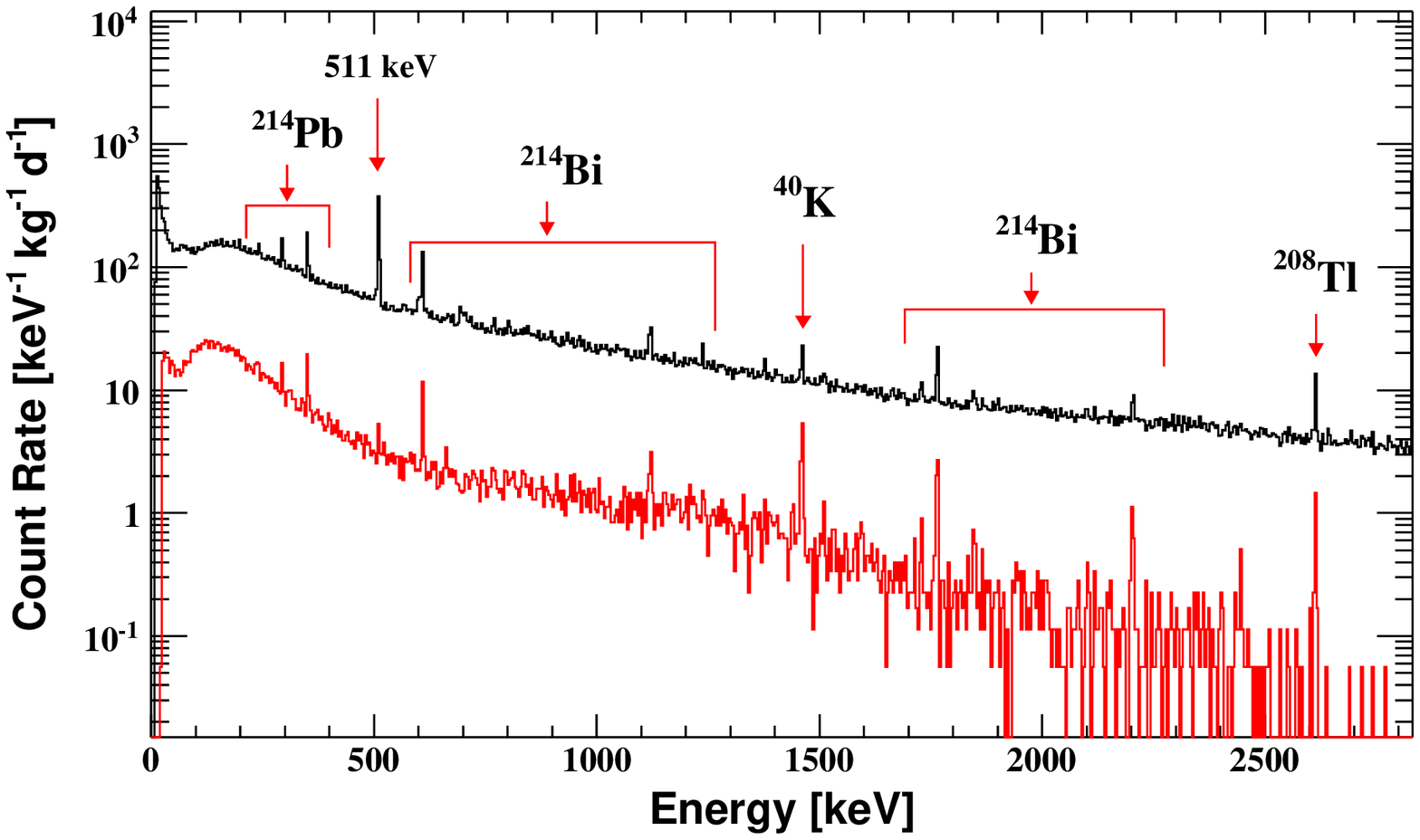}
	\caption{VT-1 surface (black/top) and underground (red/bottom) comparison. Both spectra were taken with the lead shield in place (Color online).  The underground spectrum was taken with a LN$_2$ boil-off purge in place}
	\label{fig:VT1Spectrum}
\end{figure}

The counting facility consists of two high purity germanium (HPGe) detectors specifically designed for low-background assay work. The first detector, named ``VT-1", is a commercial ORTEC LLB (very low-background) Series coaxial detector \cite{Ortec}; the high voltage filter and preamplifier are removed from the line of sight of the crystal to reduce backgrounds from radioactive contaminants that may exist in detector components.  The sample cavity is cylindrical, 41 cm (height) $\times$ 28 cm (diameter). Further specifications for VT-1 are in Table~\ref{tab:DetectorGeometries}.  A VT-1 background spectrum taken on the surface and a background spectrum taken underground is shown in Fig.~\ref{fig:VT1Spectrum}.  

The second detector, named ``{\melissa}," is a Canberra LB (low-background) coaxial detector \cite{Canberra}. {\melissa} is in a vertical orientation with a dipstick style cryostat.  The preamplifier is also removed from the cryostat, allowing for shielding to be placed in between the preamplifier and cryostat. The sample cavity is 38 cm $\times$ 38 cm $\times$ 38 cm. Further specifications for \textsc{Melissa} are in Table~\ref{tab:DetectorGeometries}.  

\begin{figure}
	\centering
	\includegraphics[width=0.5\textwidth]{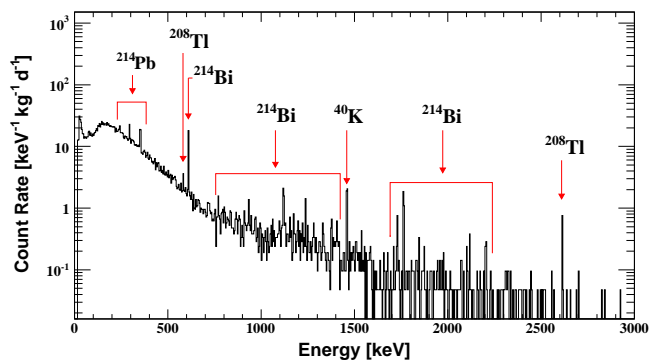}
	\caption{Background spectrum for the {\melissa} detector.  The spectrum was taken with lead shielding and a LN$_2$ boil-off purge in place (Color online).}
\label{fig:MELSpectrum}
\end{figure}

A \textsc{Melissa} background spectrum is shown in Fig.~\ref{fig:MELSpectrum}.  No active shielding is currently used for either detector. The sample cavities of both detectors are continuously purged with LN$_2$ boil-off to flush out radon. A reduction of $\sim$80\% in the activity of radon daughters was observed after introducing the LN$_2$ boil-off, however activity due to radon remains as a dominant source of background.  The shield design of the detectors has made it difficult to design a hermetically sealed radon exclusion system.

Integral count rates for these background spectra (Figs.~\ref{fig:VT1Spectrum}~and~\ref{fig:MELSpectrum}) are shown in Table~\ref{tab:CountRates}.  While direct comparisons with other underground gamma-counting facilities are difficult, integral count rates for other selected facilities can be found in \cite{Lau04,Bud07,Bud08a,Bud08b}. Although the integral count rate in the 40-2700 keV region is higher than for other detectors at a similar depth, most of our background is at low energy.    An overview of low-radioactivity background techniques and a comparison of low-level counting methods can be found in \cite{Heu95, Hult07} and references therein.
%The low energy background is likely due to $^{210}$Pb and  $^{210}$Bi Bremsstrahlung from impurities inside the cryostat.

\begin{table*}
\caption{Detector specifications.}
\begin{tabular}{l l c c} 
\hline
\hline
		&			& {\melissa}	& VT-1 \\
\hline
\multicolumn{2}{l}{Manufacturer}	& Canberra 	& ORTEC \\
\multicolumn{2}{l}{Relative Efficiency}	& 50 \%	& 35\% \\
Performance	&			&		&\\
        & FWHM at 1.33 MeV (keV)	&      1.70 & 1.80  \\
        & Threshold (keV)	&      20 & 20  \\
Shield	&			&		&\\
        & Lead thickness (cm)	&      15.2 & 10.1  \\
        & Oxygen-free high conductivity	&       &   \\        
        & (OFHC) copper thickness (cm)	&      2.54 & 0.3  \\        
Crystal (coaxial)	& 			& 		&  	\\
		& Mass (kg) &  1.1 & 0.956\\
		& Length (mm)	&	64.5	& 75.7	\\
		& Diameter (mm) &	65	& 55.8	\\
		& Hole diameter (mm) & 7.5  & 9.1 \\
		& Hole depth (mm)		&  50     & 63.2 \\
		& Outer electrode thickness (mm)       &  1.06     & 0.7 \\
		& Inner electrode thickness ($\mu$m)       &  0.3     & 0.3 \\

Cryostat	&			&		&\\
        & End cap diameter (mm) & 82 & 70 \\
        & End cap thickness (mm)      &    4.2   & 1.3 \\
        & End cap to crystal (mm)     &      5 & 4 \\
        & End cap material            &      high purity Al & Mg \\
		& IR window material       &      Al Mylar\texttrademark, Kapton & Al Mylar\texttrademark \\
		
\hline
\hline
\end{tabular}
\label{tab:DetectorGeometries}
\end{table*}

\begin{table}
\caption{Integral count rates for {\melissa} and VT-1 in [$10^3$ counts/day].}
\begin{tabular}{l c c c} 
\hline
\hline
			& Melissa	& VT-1 & VT-1 (surface)		\\
\hline
40-2700 keV 	& 7.8 	& 7.6 	& 84  \\
40-1000 keV	& 7.5 	& 6.5 	& 68  \\
1000-2700 keV & 0.27	& 0.71	& 16  \\
\hline
\hline
\end{tabular}
\label{tab:CountRates}
\end{table}

\section{Sample preparation}
To achieve the required assay sensitivity for next generation low-background experiments, unwanted radionuclides must be removed from sample surfaces to prepare for gamma-counting. Samples are prepared for assay in a cleanroom environment using ultra-pure reagents and clean plastics that have been screened for radioactivity. Depending on the sample material and required detection limit, a variety of methods can be used to treat sample surfaces. These methods include acid leaching for plastics, acid etching for metals or cleaning with ultra-pure solvents.  After cleaning, samples are bagged in nylon to prevent any recontamination of surfaces.  

\section{Analysis techniques}

\subsection{Monte Carlo simulations and efficiency calculations}

The full energy peak (FEP) detection efficiency, defined as the ratio of the number of events detected in the gamma peak to the number of events emitted from the source for a particular radioactive isotope in a sample, depends on many factors, including the crystal, cryostat, shielding and source geometries. There are currently several methods to determine the FEP detection efficiency. One is to use analytical calculations \cite{Wan95}, however this technique is limited to simple geometries and requires complex calculations. In some cases, a physical model of the sample can be created using known standards \cite{Smi09}. This process is complicated and time consuming and is of limited accuracy for complex geometries.  Another method is to use a point source calibration at representative points determined by Monte Carlo simulation. The efficiency curve generated is then corrected for absorption by a sample matrix and sample container \cite{Sae04}.

At KURF, a detailed Monte Carlo simulation for each sample is done using \textsc{MaGe} \cite{Bau06}.  \textsc{MaGe} is a \textsc{Geant4}-based \cite{Geant4,All06} simulation package maintained and developed by a joint group of the M\textsc{ajorana} \cite{MJCollaboration1,MJCollaboration2,MJCollaboration3} and G\textsc{erda}  \cite{GERDACollaboration} collaborations.  Once a detailed sample geometry has been coded into the simulation, it is uniformly doped with isotopes of interest from the $^{238}$U, $^{232}$Th, and $^{40}$K  decay chains, along with any other isotopes that may be present in the sample, e.g. $^{60}$Co.  The primary $\gamma$-rays from these decays are tracked from the emission at the source to absorption in the detector active region. By using a pure Monte Carlo simulation to determine the FEP detection efficiency, self-attenuation in the sample is accounted for and there are no limitations on source or detector configurations.  

The simulated spectrum is then convolved with the finite energy resolution of the detector. The energy resolution for both detectors has been measured as a function of $\gamma$-ray energy from 303--1836 keV using radioactive point sources. 

The peak area in the simulated spectrum is determined by fitting the peak of interest with a Gaussian and subtracting a linear background. The FEP detection efficiency, including branching ratios, can then be determined from
\begin{equation}\label{eq:eff}
\epsilon_\gamma = \frac{peak\;area}{number\;of\;events\;simulated}
\end{equation}

\subsection{Monte Carlo validation}
It is our goal to know the efficiency of any volume sample to $<$ 10\% and attribute $<$ 10\% systematic uncertainty to these efficiencies. In order to validate the Monte Carlo FEP detection efficiency calculations, point sources of well-known activity were used to baseline the simulation using a method similar to \cite{Kar02}. Once the detector geometries are well understood, the Monte Carlo can be used to simulate all relevant physical processes and accurately determine peak efficiencies. 

\begin{figure}[htp]
	\centering
	\includegraphics[width=6.03cm]{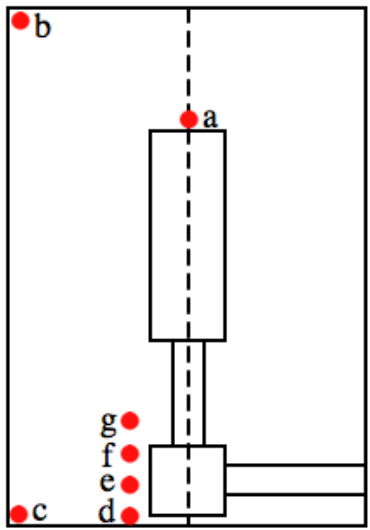}
	\caption{Point source locations in VT-1 (Color online).}
	\label{fig:VT1ptsrc}
\end{figure}

Experimental data were taken for each detector using point sources emitting $\gamma$-rays with energies ranging from 303-1836 keV. The activity of each source was known to within 3\%. To understand the effects of source placement relative to the crystal, the sources were placed in the locations shown in Fig.~\ref{fig:VT1ptsrc} for VT-1 (the same method was used for {\melissa}).  Enough experimental data were taken for each source placement to minimize uncertainties from counting statistics to $<$ 2\%.

A Monte Carlo simulation was done for each experimentally measured source location using a point source and assuming isotropic photon emission.  Simulated and experimental spectra were directly compared to determine how well the simulation agrees with the experimental spectrum.

\begin{figure} [htp]
	\centering
	\includegraphics[width=0.49\textwidth]{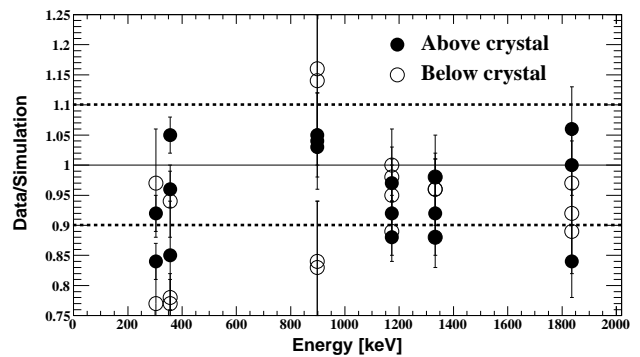}
	\caption{Experimental and simulated efficiency in the VT-1 detector. Multiple data points at each $\gamma$--ray energy correspond to a different placement of the source in the detector.}
	\label{fig:VT1_char}
\end{figure}

\begin{figure} [htp]
	\centering
	\includegraphics[width=0.49\textwidth]{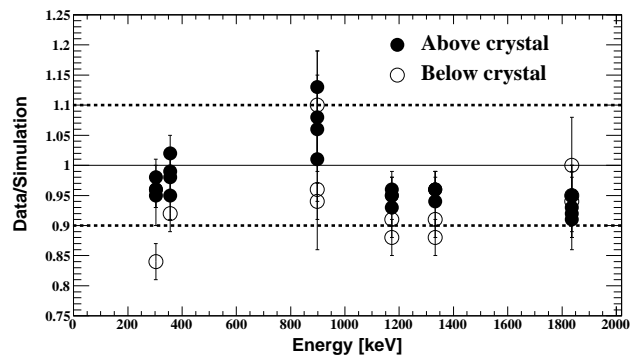}
	\caption{Experimental and simulated efficiency in the {\melissa} detector. Multiple data points at each $\gamma$--ray energy correspond to a different placement of the source in the detector.}
	\label{fig:MEL_char}
\end{figure}

The differences between the simulated efficiencies and those calculated using experimental data are shown in Figs.~\ref{fig:VT1_char}~and~\ref{fig:MEL_char}. In VT-1, the discrepancy between the simulated and experimental efficiencies is significant for regions below the crystal. The difference increases at low energies. This discrepancy is likely due to uncertainties in detector geometries below the crystal, which are not supplied by the manufacturer. The average ratio of the experimental to simulated efficiencies is 0.89 in this region. The average ratio of experimental to simulated efficiencies without considering regions below the crystal is 0.96. For {\melissa}, the ratio of the experimental to simulated efficiencies is 0.96. A constant correction factor to account for this difference is applied when determining the activity of a sample. 

\subsection{Activity Calculations}
\begin{figure}[htpb]
	  \centering
      \includegraphics[width=0.49\textwidth]{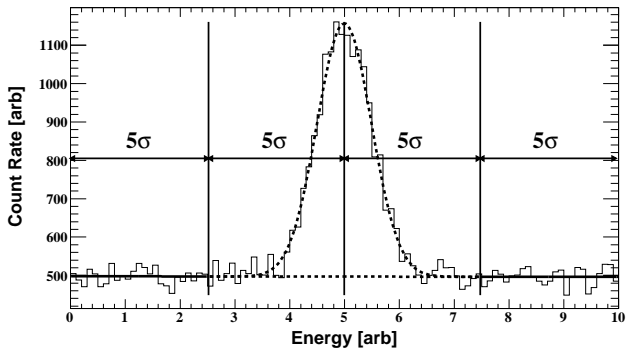}
\caption{Generic Gaussian peak as an illustration of integration limits.  The peak is fit with a linear plus a Gaussian function.  The limit of the Gaussian is $\pm5\sigma$ (dashed) and the linear background (solid) is extrapolated under the peak.} \label{fig:peakarea}
\end{figure}

With the FEP detection efficiency well understood, we can precisely determine or set an upper limit on the activity of a sample.  First, a sample spectrum is compared to the background. The background spectrum is taken close to the time of the assay to limit the effects of temporal variation in background, such as seasonal changes in radon activity. A Gaussian plus a linear background function is fit to each peak used for analysis. The fit is used to obtain the centroid of the peak and define the limits of the peak.  The peak area is then extracted by summing the counts in a $\pm 5 \sigma$ width region for both the sample and background spectra. Subtraction of the continuum is done using one of two methods: (1) Integrate the linear part of the fit extrapolated under the peak and subtract, (2) Define two regions, one 5$\sigma$ to the left and the other 5$\sigma$ to the right of the peak, average the counts in the two regions, and subtract it from the total integrated peak area.  If another peak is within 5$\sigma$ to the left or right, the background region with no peak present is used rather than the average (See Fig.~\ref{fig:peakarea} for illustration).  We prefer method (2) since this does not rely on the goodness-of-fit.   The net peak area is found by subtracting the background peak area from the corresponding sample peak.

The activity of a specific peak is then given by:
\begin{equation}\label{eq:activity}
	A_{\gamma} = \frac{net\;peak\;area\;\;[\textrm{counts/second}]}{\epsilon_{\gamma}\,m}
\end{equation}
where $\epsilon_\gamma$ is the peak efficiency, and $m$ is the mass of the sample. 

If no statistically significant peak is present, an upper limit can be placed on the activity of a sample.  The activity at a given energy is required to be less than or equal to 1.64$\times\sqrt{background \; counts}$ (90\% confidence level (C.L.)).

Once activities for individual peaks are calculated, they must be related to the overall $^{238}$U and $^{232}$Th content. In many cases secular equilibrium cannot be assumed, so activities or upper limits are considered separately for $^{238}$U (from $^{234}$Th, $^{234m}$Pa) and $^{226}$Ra (from $^{226}$Ra, $^{214}$Pb, $^{214}$Bi) in the $^{238}$U decay chain and $^{228}$Ra (from $^{228}$Ac) and $^{228}$Th (from $^{212}$Pb, $^{208}$Tl) in the $^{232}$Th decay chain. The presence of radon gas, which decays to $^{214}$Bi and $^{214}$Pb, makes the measurement of $^{226}$Ra difficult, so measures are taken to eliminate as much radon as possible inside the sample cavities through the use of LN$_2$ boil-off.

\subsection{Estimating detection limits}

A \textsc{MaGe}/\textsc{Geant4} simulation was performed to estimate the detection limits, defined as the net signal level that may be expected to lead to a detection \cite{Cur68}, for a generic sample placed in a Marinelli beaker, modeled by a Teflon{\textregistered} ring of dimensions 14.1 cm (OD) $\times$ 8.4 cm (ID) $\times$ 7.1 cm (height). The detection limits for isotopes in the $^{238}$U and $^{232}$Th decay chains and $^{40}$K were calculated for {\melissa} and VT-1 using the best-achieved background spectrum for each detector. We define a detectable signal by Eq.~(\ref{eq:sens}), where the background counts are determined from experimental background spectra and signal counts are the excess expected from the simulations of the sample. Counts within $\pm 5\sigma$ width of the signal-peak energy are included.
\begin{equation}\label{eq:sens}
  signal \; counts\geq 1.64 \times \sqrt{background \; counts}
\end{equation}
The \textsc{MaGe}/\textsc{Geant4} simulation determines the efficiency for detecting a decay, including geometric effects and branching ratios, so that the number of signal counts in a detector can be related to the source activity:
\begin{eqnarray}\label{eq:sig}
  signal \; counts= (source\;activity\;rate)\times \nonumber\\
  			  (counting\;live\;time)\times\;\epsilon_\gamma
\end{eqnarray}

The sensitivity is determined by combining Eqs.~(\ref{eq:sens}) and (\ref{eq:sig}), and solving for the detectable activity rate. This gives the sensitivity, $S$:
\begin{equation}\label{eq:sensfinal}
  S = \frac{1.64\sqrt{background\;counts}}{(counting\;live\;time)\times\epsilon_\gamma}
\end{equation}
The sensitivity results are shown in Table~\ref{tab:Sensitivity}.

\begin{centering}
\begin{table*}[hbp!]
\centering
\caption{Detector sensitivities (90\% C.L.) for {\melissa} and VT-1 for a Teflon{\textregistered} ring.}
% \begin{center}
\begin{tabular}{p{2cm} p{2.5cm} | c c | c c } 
			&							 &	\multicolumn{2}{c|}{Melissa}						& \multicolumn{2}{c}{VT-1}  \\
Energy [keV]	& Isotope (Chain)		 &  counts/day 						& Detection Limit 		&  counts/day	& Detection Limit  \\
		&					     	 &  									& [mBq/kg]		&			& [mBq/kg] \\
\hline
\hline
63			& $^{234}$Th	 ($^{238}$U)		& 81	$\pm$ 1				& 40				& 79 $\pm$ 4				& 100 \\
93			& $^{234}$Th	 ($^{238}$U)		& 96 	$\pm$2				& 60			& 121 $\pm$ 5				& 70 \\
1001			& $^{234m}$Pa ($^{238}$U) 	& 2.7	 $\pm$ 0.4			& 50				& 9 $\pm$ 1				& 230 \\
186			& $^{226}$Ra	 ($^{238}$U)		& 145 $\pm$ 3				& 30				& 121$\pm$5				& 60 \\
295			& $^{214}$Pb	 ($^{238}$U) 	& 117 $\pm$ 3				&  8				& 85 $\pm$ 4				& 10  \\
352			& $^{214}$Pb	 ($^{238}$U) 	& 114 $\pm$ 3				&  5				& 100 $\pm$5				& 10  \\
609			& $^{214}$Bi	 ($^{238}$U) 	& 59 $\pm$ 2				&  5				& 53 $\pm$ 3				& 8  \\
1120			& $^{214}$Bi	 ($^{238}$U) 	& 13 $\pm$ 1				&  8				& 21$\pm$2				& 30  \\
1764			& $^{214}$Bi	 ($^{238}$U) 	& 9.2 $\pm$ 0.7			&  8				& 18$\pm$2				& 30  \\
338			& $^{228}$Ac	 ($^{232}$Th) 	& 65 $\pm$ 2				&  10				& 48$\pm$3				& 20  \\
911			& $^{228}$Ac	 ($^{232}$Th) 	& 5.7 $\pm$ 0.7			&  3				& 13$\pm$2			         & 10  \\
238		& $^{212}$Pb ($^{232}$Th) 	& 133 $\pm$ 3				&  3				& 104$\pm$5				& 6  \\
583			& $^{208}$Tl	 ($^{232}$Th)	& 18 $\pm$ 1				&  3				& 21$\pm$2				& 8  \\
2614			& $^{208}$Tl	 ($^{232}$Th) 	& 2.9 $\pm$ 0.4			&  3				& 7$\pm$1				& 10  \\
1461			& $^{40}$K ($^{40}$K) 		& 19 $\pm$ 1				&  20				& 32$\pm$3			         & 50 \\
\hline
\end{tabular}
\\
%*Denotes best detection limit to $^{238}$U \\
%**Denotes best detection limit to $^{232}$Th
\label{tab:Sensitivity}
\end{table*}
\end{centering}

\section{Assay results}
Table~\ref{tab:AssayResults} shows the sample assay results to date since May 2008. Samples are listed in the order that they were assayed. Activity limits for later assays were improved as a result of progress made with background reduction and detector performance. Disequilibrium in natural unclosed systems, such as plants, soil and rock, is common as observed in the Table Mountain rock sample. For the sample of aluminum stock flange coupling, disequilibrium was observed in the $^{238}$U and $^{232}$Th decay chains. The decrease in activity of the signature $^{226}$Ra daughters have been observed in previous analyses of both pure and aluminum alloys. This can be explained through differences in the chemistries of U, Th, Ra and Pb, as could occur through the steps required to recover aluminum from its ore \cite{Smi08}. For lead samples, $^{210}$Pb ($T_{1/2} = 22.3 \cdot y$), activity was measured separately since it is not expected to be in equilibrium with the rest of the $^{238}$U chain. The lead sample from Sullivan Metals had less than 2.5 Bq/kg of $^{210}$Pb activity.  Lead bricks of unknown origin acquired from the University of Washington showed no measurable $^{210}$Pb signatures, but an upper limit of 10 Bq/kg was placed on the $^{210}$Pb activity.  

\begin{table*}
\caption{KURF assay results. $^{234}$Th - activity measured in the $^{238}$U decay series from $^{234}$Th and $^{234m}$Pa. $^{226}$Ra - activity measured in the $^{238}$U decay series from $^{226}$Ra, $^{214}$Pb and $^{214}$Bi. $^{228}$Ra - activity measured in the $^{232}$Th decay series from $^{228}$Ac. $^{228}$Th - activity measured in the $^{232}$Th decay series from $^{212}$Pb and $^{208}$Tl. Activities from $^{208}$Tl are divided by the branching ratio (35.94\%). Not measured = NM.}
\begin{tabular}{|p{3cm} | l | l | l | l | l | l | l |} 
\hline
Sample                              					& Detector                & $^{234}$Th [Bq/kg] 	     & $^{226}$Ra [Bq/kg] & $^{228}$Ra [Bq/kg]  & $^{228}$Th [Bq/kg]   		& $^{40}$K [Bq/kg] 		&$^{60}$Co [Bq/kg]\\
\hline
Table Mountain rock (latite)     					& {\melissa}  	       &          NM                & 100$\pm$40   & 100$\pm$40        & 270$\pm$120  	 & 790$\pm$320	&NM\\ \hline
Table Mountain rock (latite)                				&VT-1                        &           NM               & 100$\pm$40   & 100$\pm$40        & 300$\pm$120  	& 730$\pm$290	&NM\\ \hline
Superinsulation panels                       			& {\melissa}                 &         NM                  & 3.0$\pm$1.2   &                NM                & 0.09$\pm$0.03	 & 0.9$\pm$0.4		&NM\\ \hline
Aluminum stock flange coupling      			& {\melissa}                 &7.1$\pm$2.3    & 1.5$\pm$0.4   &  $<$0.05                 & 1.5$\pm$0.4           & $<$0.2			&NM\\ \hline
PMT base electronic components     			& {\melissa}                 &$<$4                  & 1.5$\pm$1.0   &  0.8$\pm$0.6       & 0.6$\pm$0.4     	& 3.3$\pm$1.9		&NM\\ \hline
PMT base electronic components     			& VT-1                      &$<$2               & 1.1$\pm$0.7   &   0.8$\pm$0.4        & 0.6$\pm$0.3     	& 3.6$\pm$1.8		&NM\\ \hline
Zeolite molecular seive              				& {\melissa}                 &5.8$\pm$1.2    & 8.2$\pm$0.8   &   9.6$\pm$0.6       & 10.5$\pm$0.6   	& 4.4$\pm$0.5		&NM\\ \hline
Great Stuff{\texttrademark} foam insulation 	& {\melissa}                &             NM              & $<$0.2             &                  NM                & $<$0.2                	&  $<$0.3			&NM\\ \hline
Axon Picocoax{\textregistered}          			& VT-1	               & $<$0.7              & $<$0.2          & 0.060$\pm$0.020 & 0.055$\pm$0.010   &700$\pm$200          & $<$0.01\\ \hline
Sullivan lead bricks						& {\melissa}		     & $<$0.01		        &$<$0.002		& $<$0.0005               &$<$0.0004               &$<$0.003                  &NM\\ \hline
University of Washington lead bricks		&{\melissa}		   & $<$0.01         &$<$0.003         &$<$0.001               &$<$0.0004                 &$<$0.003                  & NM\\  \hline
PEEK plastic							& VT-1		   & $<$0.2		& $<$0.04         & $<$0.04 		& $<$0.03		&$<$0.14		& $<$0.008 \\ 
\hline
\end{tabular}
\label{tab:AssayResults}
\end{table*}

\section{Conclusions \& Remarks}
A gamma-counting facility has been commissioned at KURF.  The background signal rates for the {\melissa} and VT-1 detectors have been pushed to low levels.   This was accomplished by building the facility in an underground location and using passive shielding, radio-pure detector components, and radon mitigation techniques.  We have successfully demonstrated the analysis procedures and assay sensitivities required for screening materials for the next generation of low-background experiment.

\section{Acknowledgements}
This worked was primarily support by NSF grant \# PHY 0705014 and DOE-HEP grant \# DE-FG02-07ER41484. Additional support was provided by DOE grant numbers DE-FG02-97ER4104 and DE-FG02-97ER41033. We acknowledge Lhoist North America, especially Mark Luxbacher, for providing us access to the underground site and logistical support. We also acknowledge the logistical support and help with remote operations from Mary Kidd, S. Derek Roundtree, Werner Tornow, and the TUNL technical staff.

\bibliography{kimb_lbc}

\end{document}